\documentclass[11pt]{article}

\usepackage[margin=1in]{geometry}
\usepackage[T1]{fontenc}
\usepackage[utf8]{inputenc}
\usepackage{microtype}
\usepackage{graphicx}
\usepackage{float}
\usepackage{booktabs}
\usepackage{tabularx}
\usepackage{array}
\usepackage{amsmath,amssymb}
\usepackage{caption}
\usepackage{subcaption}
\usepackage{xurl}
\usepackage{titlesec}
\usepackage[hidelinks]{hyperref}
\usepackage[backend=biber,style=vancouver,sorting=none]{biblatex}
\usepackage[section]{placeins}

\newcommand{\headingragged}{\raggedright\hyphenpenalty=10000\exhyphenpenalty=10000}
\titleformat{\section}{\normalfont\Large\bfseries\headingragged}{\thesection}{1em}{}
\titleformat{\subsection}{\normalfont\large\bfseries\headingragged}{\thesubsection}{1em}{}
\titleformat{\subsubsection}{\normalfont\normalsize\bfseries\headingragged}{\thesubsubsection}{1em}{}

\graphicspath{{figures/}}

\title{\textbf{Clear Mind: \\
Meditation and the Brain's Signal-to-Noise Ratio}}
\author{
\begin{tabular}{c}
Ruben E. Laukkonen$^{1,2,3,4}$\\[0.75em]
\small $^{1}$Flourishing Intelligence Program, Linacre College, University of Oxford, Oxford, UK\\
\small $^{2}$Centre for Eudaimonia and Human Flourishing, Linacre College, University of Oxford, Oxford, UK\\
\small $^{3}$Department of Psychiatry, University of Oxford, Oxford, UK\\
\small $^{4}$LIFE, London, UK
\end{tabular}
}
\date{}

\begin{document}

\maketitle

\begin{abstract}
Meditation is quintessentially connected with a clear mind. This paper proposes that diverse findings in the science of meditation can be mapped onto a single, empirically tractable construct: the brain's functional signal-to-noise ratio (f-SNR). Here, ‘signal’ denotes neural variance that tracks the goal-relevant causes of sensory input, while ‘noise’ denotes residual activity including  irrelevant endogenous fluctuations. Mechanistically, meditation boosts f-SNR through two primary operations: selectively enhancing signal and "decluttering" noise. Deepening practice is further proposed to enhance f-SNR via reducing self-referential filtering and shifting global neural activity toward a critical regime—a thermodynamically efficient state that maximizes information transmission and dynamic range. This framework has a strong existing evidence-base and is readily falsifiable using metrics such as neural variability quenching, mutual information, and multivariate decoding. Notably, our first test of the theory demonstrates that multiple indicators of neural SNR increase alongside meditation depth. The f-SNR account offers a transdiagnostic explanation for the efficacy of meditation across a multitude of psychopathologies associated with low SNR states. The theory also has intriguing implications for emerging technology: meditation may improve Brain-Computer Interfaces (BCIs) by making the brain easier to read.
\end{abstract}

\section{Introduction}

\begin{quote}\itshape “Breathing in, I see myself as still water. Breathing out, I reflect things as they are.”\end{quote}

\begin{flushright}--- Thich Nhat Hanh \cite{nhathanh1992touchingpeace}\end{flushright}

There are scarce scientific generalisations regarding the diverse effects of meditation \cite{VanDam2018,Sparby2025,Ehmann2025}. Yet, for millennia contemplatives have reported a clearer mind \cite{KabatZinn1994,Metzinger2024}. This phenomenological touchstone, often described as lucidity, vividness, or stillness \cite{Metzinger2024,Dunne2013,Yang2024}, may indicate a rare point of convergence. This paper proposes that many findings in the science of meditation can be mapped onto a neurobiologically grounded and empirically tractable construct: the brain's functional signal-to-noise ratio (f-SNR).

The brain must model the world to act adaptively in it \cite{Friston2010}. The degree to which this ‘world model’ aligns with the information geometry of the incoming information can be modeled through the f-SNR between biological activation patterns and task-relevant sensory data. To put it intuitively: By bringing the mind into a continuous resonance with incoming ‘present moment’ data, meditation may naturally improve the degree to which world modelling is titrated to the ongoing stream of input.

Mechanistically, meditation modulates the f-SNR via two channels: decluttering, and enhancing—i.e., reducing the task-irrelevant activity and boosting the relevant, respectively. This is consistent with a wide array of practices that seek to “let go” some content of experience (e.g., neurotic thinking) in favour of a chosen goal-state (e.g., sustained present-moment awareness) \cite{Bishop2004,Dahl2015,Laukkonen2021}. But it is also consistent with advanced meditation, which resembles a rewiring of deeply engrained habits that betray our ability to see the world and ourselves clearly, such as excessive reactivity or self-centred thinking.

Complementing this, emerging evidence indicates that meditation tunes neural activity toward criticality \cite{Pascarella2025,Mago2025,Vohryzek2025}—a balanced regime between excessive synchrony and fragmentation where information transfer and dynamic range are maximal \cite{Cocchi2017}. Criticality may further boost SNR by permitting a broader dynamic range (appropriate for a complex world in flux) and sensitivity to new data, making the mind a more responsive “mirror”.

The argument here is not that meditation always succeeds in improving the brain’s SNR, or that all practices and arising experiences can be explained by the theory. But that SNR is a useful and parsimonious explanation that captures meaningful variance across styles of practice throughout a long and complicated meditation journey. It is also straightforward to test and could eventually function as a yardstick for expertise; the absence of which is a fundamental problem in meditation research \cite{VanDam2018,Sparby2025,Ehmann2025}.

The SNR view is also practically important. If we understand that one of the central goals of meditation is to improve the brain’s SNR, then we can refine practices and technologies to efficiently achieve mental clarity. We may also find that some psychopathologies (e.g., those underlying delusions, obsessions, excessive rumination, and other maladaptive beliefs) are significantly underwritten by low neural SNR to the present-moment unfolding sensory inputs; thwarting appropriate updates to priors \cite{NolenHoeksema2008,Gillan2014,Kapur2003,Sterzer2018}.

The paper begins with a basic introduction to the notion of SNR in the context of a brain trying to model the causes of sensory input. It then describes how the f-SNR view addresses multiple explanatory targets of meditation, and the diverse ways that different meditation styles selectively tune diverse channels of signal and noise. A concise review of converging evidence is then provided alongside a toolkit for falsification. The f-SNR framework is then extended to information theoretic perspectives, including criticality and dynamical systems theory—ultimately showing how meditation can make the brain more metabolically efficient and responsive to ongoing causes of sensory input. I review key pieces of existing evidence and conclude with implications for emerging technologies.

\section{The Geometry of Shared Variance Between Brain and World}

\subsection{Instrumental vs. Functional SNR}

In neuroscience, SNR is the gold standard for determining the fidelity of a neuroimaging device \cite{Epistatou2020,Welvaert2013}. It can be calculated by dividing the mean amplitude of a signal of interest by the standard deviation of the background physiological or instrumental noise. In electroencephalography (EEG), magnetoencephalography (MEG), or functional magnetic resonance imaging (fMRI), this typically involves comparing the power of a stimulus-evoked or task-related response to the variability of baseline activity or scanner noise in different ways \cite{Parks2016,GonzalezMoreno2014,Welvaert2013}. The resulting ratio reflects how reliably the system can capture real neural fluctuations. Let us call this instrumental SNR.

Just as we can compute SNR for an imaging device that is trying to capture the activity of the brain, we can also frame the brain as a ‘device’ trying to capture the activity of the world for adaptive, epistemic, action \cite{Friston2009,Friston2015}. In other words, if we assume that organisms are engaged in continuously modelling the world for survival, then one way to quantify the fidelity of this internal model is through the shared information between the true causes of sensory input and measurable neurobiological activity: namely, functional SNR (f-SNR).

Technically, a high f-SNR indicates that, over time, neural activity selectively enhances representational states that reduce uncertainty about relevant causes, while suppressing fluctuations that do not improve inference. This aligns with information-theoretic formulations of mutual information, where shared information between two variables can be understood as the reduction of uncertainty in one variable given knowledge of another \cite{Borst1999}.

A simple expression of this decomposition for a neural response r to a task-relevant variable z follows from the law of total variance:

\[
\operatorname{Var}(r) = \underbrace{\operatorname{Var}_{z}\bigl(\mathbb{E}[r \mid z]\bigr)}_{\text{signal}} + \underbrace{\mathbb{E}_{z}\bigl(\operatorname{Var}(r \mid z)\bigr)}_{\text{noise}}
\]

This identity (the law of total variance) decomposes the total variability of a neural response r into two additive components with respect to a task variable z. The first term (“signal”) measures how much the mean response changes across conditions (e.g., average firing rate differences between two visual stimuli), while the second (“noise”) quantifies how variable responses are within each condition (e.g., trial-to-trial fluctuations to the same image) \cite{Weber2021}. Adding these components reconstructs the total variance observed across all trials, providing a principled decomposition of uncertainty in neural coding and permitting us to compute the SNR (i.e., S/N).

Conceptually, f-SNR simply concerns the informational clarity of the brain’s own computations and empirically can be inferred in many ways (discussed below). As we will see later, f-SNR also has implications for instrumental SNR, since clearer and more stable neural signals are easier to measure, decode, and reproduce \cite{Gallego2020,Pancholi2023,Lycke2023}—effectively making the brain a cleaner source for the instrument itself (such as brain-computer interfaces; BCIs). Figure 1 provides visual illustration of what might broadly be expected across neuroimaging devices if the f-SNR theory holds true (more approaches described later).

\begin{figure}[htbp]
  \centering
  \caption{Simulated data illustrating meditation’s expected effects on the brain’s f-SNR}
  \label{fig:1}
  \includegraphics[width=0.95\textwidth]{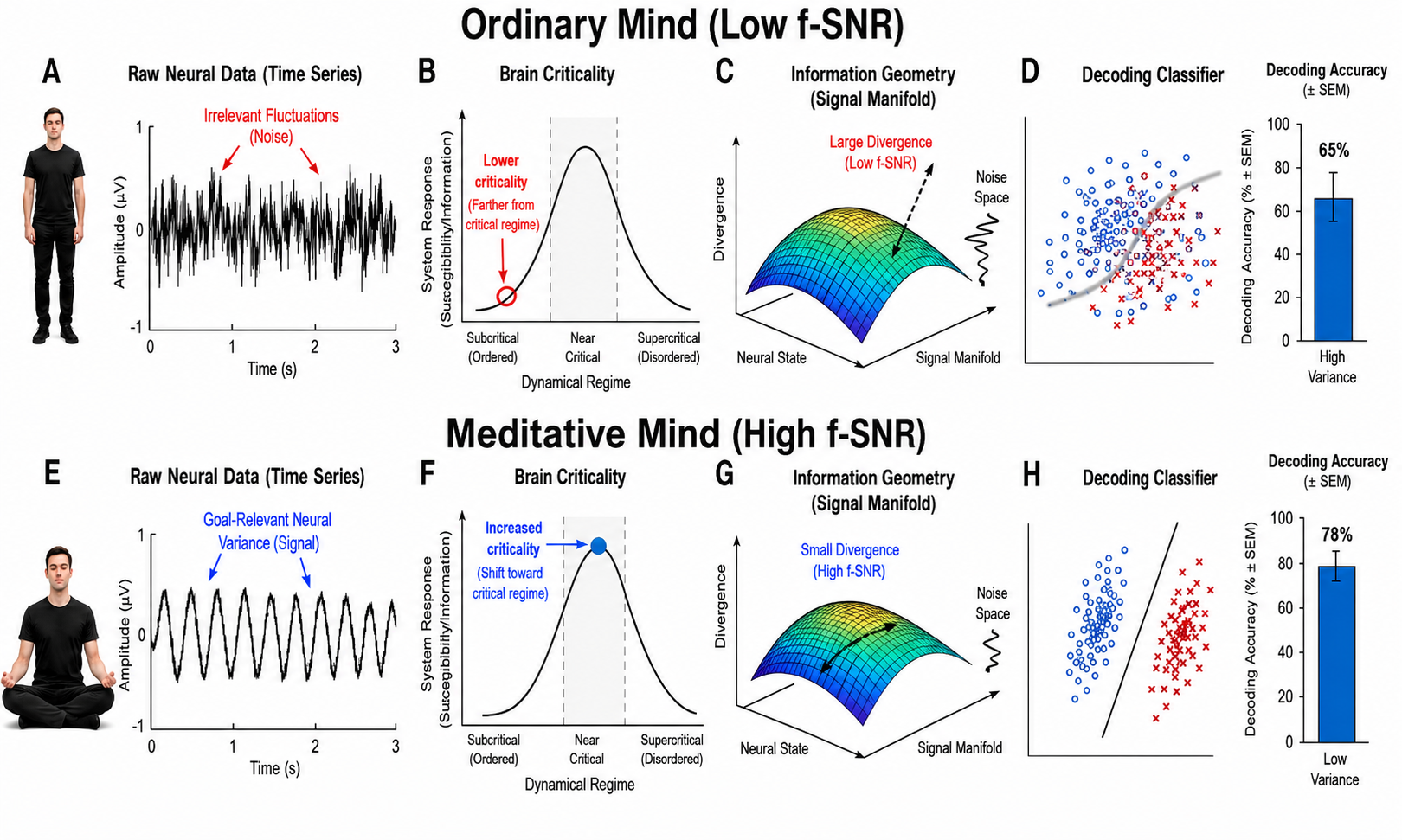}
  \begin{minipage}{0.95\textwidth}
  \small Note. This figure illustrates the hypothesis that meditation shifts the brain from a state of low f-SNR to a state of high f-SNR using simulated data. (A \& E): Raw Neural Data (EEG Time Series): Panel A shows typical neural activity characterized by high amplitude "irrelevant fluctuations" representing noise. Panel E shows the meditative state, where "goal-relevant neural variance" (signal) is more prominent and rhythmic. (B \& F) Brain criticality: The ordinary mind (B) shows a sub-critical neural regime. The meditative mind (F) sits closer to criticality, indicative of improved information processing efficiency. (C \& G) Information Geometry: These 3D visualizations depict the trajectory of neural states relative to an ideal signal manifold (blue grid surface). The ordinary mind (C) shows large divergence, with the neural state moving erratically into noise space away from the manifold. The meditative mind (G) shows small divergence, with the neural state adhering closely to the trajectory of the signal manifold. (D \& H) Decoding Classifier \& Accuracy: The final columns demonstrate the practical implication of improved SNR. The decoding classifier plots show that neural states in the meditative mind are more distinct and linearly separable compared to the blurred, complex boundary in the ordinary mind. NB: These results intentionally exemplify idealized results unlikely to portray "messy" human data, cf. Figure 4 to see results from our first experiment testing the present theory. 
  \end{minipage}
\end{figure}

\section{How Meditation Clears the RAM}

As noted at the outset, there is an intuitive connection between the phenomenology of a “clear mind” and the idea that the brain might be operating at a high f-SNR. Indeed, the sense that meditation improves the clarity, vividness, and freshness of one’s mind is a ubiquitous notion in contemplative circles. For example, in Metzinger’s book “Elephant and the Blind”, clarity emerged as a recurring feature in the >500 phenomenological accounts of meditators, alongside the related notion of “luminosity” \cite{Metzinger2024}. Despite this, no known neural mechanism currently maps onto the experience of mental clarity. Here I aim to provide a general sense for how clarity, indexed via f-SNR, can provide a parsimonious way to understand outcomes and practices associated with meditation.

\subsection{Mindfulness}

Much of the neuroscientific work on meditation is focused on Buddhist styles. Among these, the most well-studied across disciplines, from neuroscience to clinical psychology, is mindfulness. Mindfulness is the practice of observing and recognizing the flow of contents and states of mind, without judgment, while directing attention to the present moment \cite{KabatZinn1994,KabatZinn2003,Bishop2004}. In f-SNR terms, mindfulness can be construed as an exercise that declutters endogenously generated distractions, such as mind-wandering, and amplifies the signal of ongoing sensory inputs (see Figure 1). The non-judgmental component also plays a critical functional role: it prevents the system from adding redundant evaluative or self-referential “noise” and encourages learning or insights, for example via restructuring of priors, to take place as one reflects on how experience changes over time \cite{Bishop2004,Creswell2017}.

In the language of computational neuroscience, this is equivalent to tuning up the gain on input and feed-forward processing, reducing unnecessary top-down processing, and efficiently modeling stimuli at the right “level” of hierarchical engagement or degree of abstraction. This interpretation is consistent with hierarchical accounts of cortical organization, in which sensory inputs are progressively transformed across cortical processing stages into increasingly abstract representations \cite{Felleman1991}, as well as with deep-learning-inspired models of sensory cortex that map lower-level features into higher-level object and conceptual representations \cite{Yamins2016}. This means that mindfulness may improve f-SNR specifically by allowing agents to regulate, and learn, the selective engagement of hierarchical processing (cf. Figure 2). For example, while hearing someone share a traumatic event, it may not be appropriate to “problem solve” with highly abstract notions. It may be better to engage with interoceptive feelings and openly and unconditionally listen to the incoming words and feelings conveyed, while adapting dynamically.

\begin{figure}[htbp]
  \centering
  \caption{How mindfulness trains adaptive and selective engagement of hierarchical processing}
  \label{fig:2}
  \includegraphics[width=0.95\textwidth]{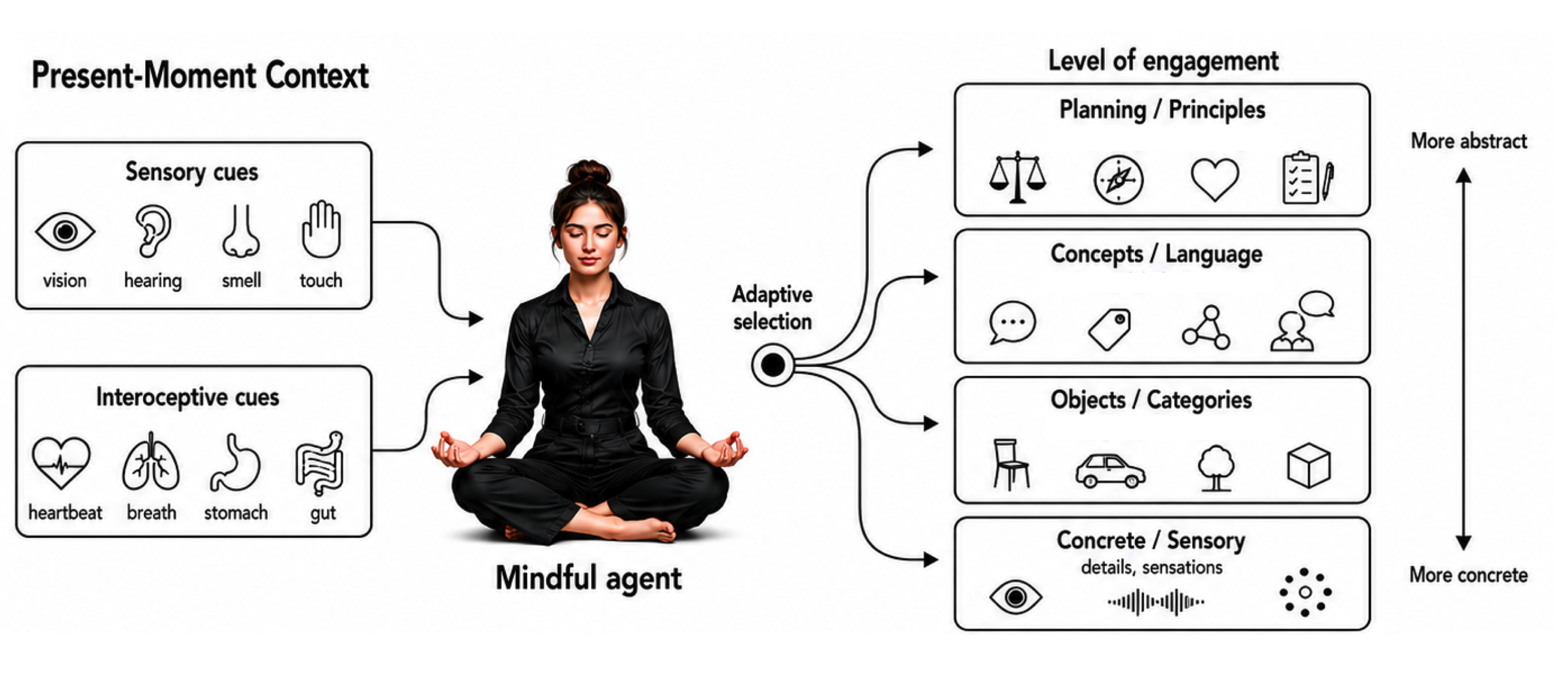}
  \begin{minipage}{0.95\textwidth}
  \small Note. The schematic depicts mindfulness as a regulatory process that can flexibly tune engagement with hierarchical processing. The mindful agent can orient processing along an abstract-to-concrete continuum. Depending on the situation, engagement may be directed toward abstract principles and planning, objects and categories, or concrete sensory details. 
  \end{minipage}
\end{figure}

\subsection{A Spectrum of Practices}

When placed on a spectrum from beginner to advanced \cite{Sparby2025,Ehmann2025,Laukkonen2021,Dahl2015}, meditation often begins with sustained attention on some sensory or mental object at the exclusion of any arising distractions, such as thoughts or fluctuations of attention (termed focused attention). Other intermediate practices maintain an open (or partially open) field of awareness, allowing various mental and sensory events to arise and pass, while “letting go” any sustained distractions (e.g., mind-wandering). These practices are regularly called open monitoring. Vipassanā or insight meditation can be a variation of focused attention or open monitoring, if it includes careful investigation of the various changes in content and states. Other subgroups of ‘advanced’ practices include non-dual meditations (indicating the direct recognition of an awareness that transcends or underlies the subject-object duality, \cite{Metzinger2024,Dunne2013}) as well as a spectrum of styles and states broadly described as jhanas (progressively deep states of tranquillity or absorption, \cite{Yang2024}). Other important practices are explicitly “constructive”, aiming to generate states of loving kindness, compassion, or identity with imaginal structures \cite{Dahl2015}. This list is far from exhaustive, but provides a set of pragmatic explanatory targets.

For the purposes of the f-SNR framework, these practices can be framed in terms of their effects on specific channels of signal and noise; wherein advanced practices address deeper sources of noise (e.g., associated with self-processing). Of course, precisely what counts as signal and noise is relative to tradition (particularly in defining the amplified channel). But there is also ground truth: The shared variance between brain and world. As practice deepens, we should see increases in the sustained and generalisable f-SNR of the brain (cf. Table 1). In some instances, we might be surprised to find that focused attention on a particular stimulus improves f-SNR compared to non-dual meditation or open monitoring. But the broader prediction here is that in temporally extended, complex and immersive environments (reminiscent of day-to-day life), the amount of mutual information between brain, body, and world, would be higher in the case of an advanced practitioner enacting advanced states.

\begin{table}[htbp]
  \centering
  \footnotesize
  \caption{Types of noise, and types of signal, organised by meditation style.}
  \label{tab:1}
  \begin{tabularx}{\textwidth}{>{\raggedright\arraybackslash}X>{\raggedright\arraybackslash}X>{\raggedright\arraybackslash}X}
    \toprule
    Practice & Noise reduced (Declutter channel) & Signal increased (Amplify channel) \\
    \midrule
    Focused Attention & Mind-wandering, distractor capture, evaluative thought & Sensory precision on the chosen object; top-down stability and conflict monitoring \\
    Open Monitoring & Reactive elaboration, cognitive fusion, and identification with mental contents & Meta-awareness; flexible attention allocation across sensory and interoceptive streams \\
    Vipassanā (Insight) & Conceptual overlay and revision of miscalibrated priors & Fine-grained feature tracking; updating and simplification of perceptual and cognitive models \\
    Non-dual Awareness & Self-referential filtering; hierarchical bias toward the “self-model” & Integrated coupling between perception and cognition; more accurate/flexible self-model \\
    Loving-Kindness & Threat bias, self-focused rumination, affective defensiveness & Prosocial and affiliative signals; approach motivation; affective coherence and warmth \\
    \bottomrule
  \end{tabularx}
\end{table}

\subsection{Posture as f-SNR error correction}

While much of meditation research focuses on internal mental acts, the physical posture itself may function as an essential biological "hardware" component for regulating f-SNR. In traditions utilizing rigorous postures like Padmasana (Lotus) or Seiza (used in Zazen), the body may act as a kind of closed-loop "neurofeedback" system. The vertical alignment of the spine against gravity requires a precise, low-energy tonic engagement \cite{Horak2006,Peterka2002}. When the mind wanders (increasing neural noise), this delicate muscular tone is lost, and the posture collapses into slumping. This physical collapse generates an immediate error signal via proprioceptive and vestibular afferents, alerting the practitioner to the loss of signal fidelity \cite{Proske2012,Angelaki2008}. Thus, the posture acts as a physical constraint mechanism that externalizes the state of the brain’s SNR. By committing to a posture that requires vigilance to maintain, the meditator creates a loop where high f-SNR, or alertness to interoceptive input, supports the posture, and the posture, in turn, scaffolds high f-SNR by punishing “noise”, such as distraction and mind-wandering, with physical instability (cf. Figure 3).

\begin{figure}[htbp]
  \centering
  \caption{Bodily posture as an embodied sensorimotor error-correction loop}
  \label{fig:3}
  \includegraphics[width=0.95\textwidth]{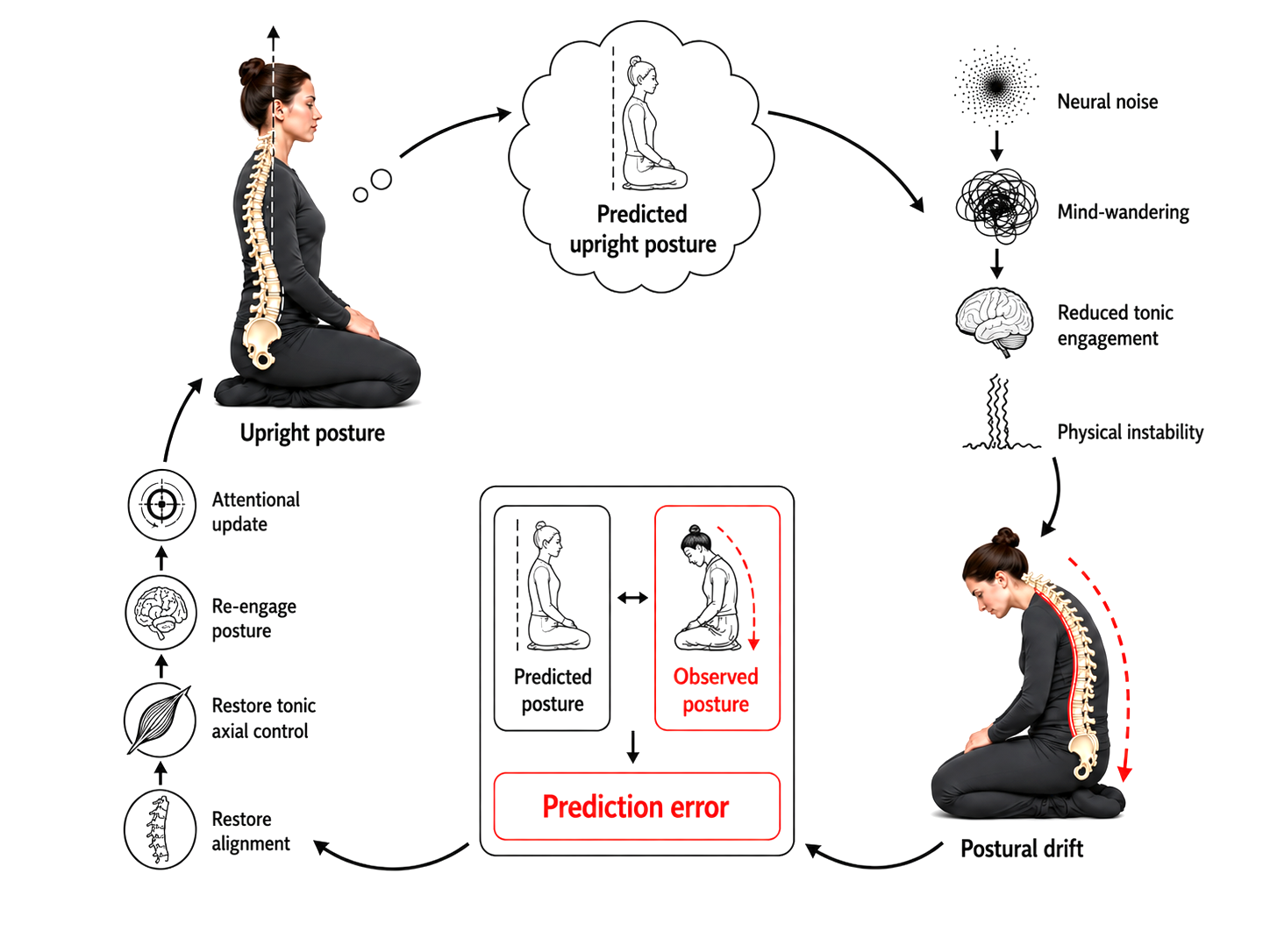}
  \begin{minipage}{0.95\textwidth}
  \small Note. A schematic of a practitioner in a traditional posture (e.g., Zazen). The upright spine represents the optimal "Signal State," requiring continuous, low-level tonic activation from the brain. As the brain’s f-SNR drops (e.g., during mind-wandering or lethargy), the resulting physical slump triggers vestibular and proprioceptive sensors, acting as an error signal (feedback). This prompts the practitioner to "Declutter" the distraction and re-engage the posture, thereby boosting the global f-SNR.
  \end{minipage}
\end{figure}

\section{Converging evidence supporting the f-SNR framework in meditation}

In service of a concise paper, a full literature review is beyond scope; yet, existing empirical work already coheres well with the f-SNR framework.

\textbf{Stable and reliable neural activity. }A brain with a high f-SNR is not only clearer but also more stable and reliable. This prediction is supported by evidence that the brain activity of long-term meditators is less variable over time. Studies using massive time-series feature extraction on EEG data have found that meditators exhibit higher temporal stationarity and stability compared to non-meditators, a finding thought to underpin greater attentional stability \cite{Bailey2024}. Lutz et al. \cite{Lutz2009} also provide a direct longitudinal demonstration of increased attentional stability consistent with higher f-SNR. After three months of intensive meditation training, participants showed reduced trial-to-trial reaction time variability on a dichotic listening task and increased theta-band phase consistency of EEG responses over anterior sites to target tones. Individuals who increased neural response consistency the most also reduced behavioural variability the most, linking improved performance to reduced neural “jitter.” Notably, the training also reduced variability in low-frequency neural processing regardless of whether the deviant tone was attended or unattended, consistent with a more globally stable processing regime

\textbf{Efficient information processing.} A high-SNR brain should also allocate its limited computational resources more efficiently. Put differently, the mechanism of action underlying improved resource allocation in meditation may be driven by boosts in f-SNR. This view is supported by a wide range of studies. A landmark study by Slagter et al. \cite{Slagter2007} examined the "attentional blink," a phenomenon where observers often fail to detect a second target (T2) if it appears in close succession to a first (T1). They found that after three months of intensive meditation training, participants showed a significantly reduced P3b brain potential in response to T1. The P3b is thought to index the allocation of attentional resources. This reduction in resource allocation to T1 was correlated with a smaller attentional blink—that is, better performance in detecting T2. With regard to the f-SNR theory: by processing T1 more efficiently (i.e., with a clearer signal requiring fewer resources), there are more resources available to process T2.

Similar findings of enhanced efficiency come from longitudinal studies using the Stroop task, a classic measure of cognitive control. Moore et al. \cite{Moore2012} found that 16 weeks of brief, daily mindfulness practice leads to an enhanced N2 component (indicating improved conflict monitoring) but a reduced P3 component, signifying more efficient, less resource-intensive processing of incongruent stimuli. Together, these ERP studies demonstrate that meditation trains the brain to achieve better cognitive outcomes with a more economical use of neural resources—hallmarks of high f-SNR.

Schöne et al. \cite{Schone2018} provide further longitudinal electrophysiological evidence of efficient information processing. After eight weeks of mindful breath-awareness practice (with progressive muscle relaxation as an active control), participants improved multiple-object tracking performance while showing reduced steady-state visually evoked potential amplitudes. This pattern is consistent with increased functional signal-to-noise ratio: better discrimination and maintenance of task-relevant information with less stimulus-locked neural expenditure, plausibly via more selective precision allocation (enhancement) and reduced engagement with task-irrelevant fluctuations (decluttering). MacLean et al. \cite{MacLean2010} extended this logic to vigilance limits wherein meditation improved visual discrimination and attenuated the vigilance decrement during sustained attention. The authors explicitly interpret the effects through a resource model: improved perceptual sensitivity (i.e., f-SNR) reduces the resource demand of target discrimination, making it easier to maintain voluntary attention

\textbf{Selective decluttering and signal enhancement. }Further causal evidence comes from studies using combined transcranial magnetic stimulation and EEG (TMS-EEG). One recent study \cite{humble2025tmseeg} found that experienced meditators have a significantly larger P60/N100 ratio in response to TMS pulses delivered to the dorsolateral prefrontal cortex (DLPFC). This suggests an altered balance of cortical excitation and inhibition, potentially reflecting greater inhibitory control that renders the cortex more stable and less susceptible to noisy, irrelevant perturbations. Indeed, an increased P60/N100 ratio is linked to improved working memory, potentially capturing improved efficiency of information processing \cite{Chung2019}.

Oddball paradigms provide convergent evidence that meditation can both declutter distractor processing and, when task demands require it, enhance target-relevant signalling. In a passive three-stimulus auditory oddball, Cahn and Polich \cite{Cahn2009} found that Vipassana meditation reduced distractor-evoked indices of involuntary orienting (notably a reduced P3a to distracters, alongside reduced frontal distractor N1 and reduced P2), consistent with diminished automatic reactivity to salient but irrelevant events. Time–frequency work in a similar oddball context suggests a shift toward “…enhanced perceptual clarity and decreased automated reactivity”, as evidenced by altered distractor-related dynamics (e.g., reduced delta power to distracters) and enhanced phase-synchrony markers to background stimuli, plus expertise-linked early gamma responses \cite{Cahn2013}. When the oddball includes an explicit target discrimination demand, Vipassana practice can instead increase target-related updating/engagement markers: Delgado-Pastor et al. \cite{DelgadoPastor2013} reported larger P3b amplitudes to targets after a meditation period relative to pre-meditation and a matched control condition. By contrast, evidence that meditation reliably modulates early deviance-detection signals such as mismatch negativity (MMN) remains mixed; a higher-powered replication attempt reported null MMN effects of focused-attention and open-monitoring meditation with Bayesian evidence against state/expertise interactions \cite{Fucci2022}.

Recent work clarifies this picture by showing that MMN may be selectively enhanced in highly absorbed practice states: during a 10-day retreat, experienced practitioners exhibited larger frontocentral MMN during jhana (absorptive meditation) than during standard mindfulness of breathing, alongside shifts in EEG dynamical markers consistent with a metastable, near-critical regime \cite{Mago2025}. This pattern is consistent with an f-SNR account in which advanced absorption reduces elaborative reactivity and sensory content while preserving (or even sharpening) perturbational sensitivity to unexpected events, yielding clearer deviance signalling against a quieter background.

Recent resting-state EEG evidence provides a particularly relevant test of the amplification or enhancement channel because it separates oscillatory activity from the aperiodic 1/f background. McQueen et al. \cite{McQueen2024} compared experienced meditators with matched non-meditators using methods that distinguished oscillatory activity from the aperiodic background, allowing oscillatory theta, alpha, beta, and gamma activity to be estimated after controlling for non-oscillatory 1/f activity. Meditators showed higher theta, alpha, and gamma oscillatory power, with differences driven substantially by altered scalp distributions rather than only global power increases. In f-SNR terms, this suggests that meditation experience is associated not merely with more neural activity, but with a more structured distribution of band-limited activity. However, because meditators did not differ from non-meditators in 1/f slope or intercept, this finding should be interpreted as evidence for altered oscillatory organization rather than as evidence that meditation globally reduces aperiodic activity.

\textbf{Clearer state-dependent multivariate patterns.} If meditation enhances f-SNR, then internal mental states should become more reliably distinguishable in neural activity, effectively increasing their decodability with multivariate methods. Consistent with this, Weng et al. \cite{Weng2020} used participant-specific MVPA on fMRI to classify breath-focused attention, mind-wandering, and self-referential processing above chance, and decoding generalized to an independent meditation run. Crucially, the decodability of the mental states showed stronger generalizability in experienced practitioners, indicative of clearer state-dependent multivariate neural activity. Decoding accuracy (cf. Table 2 and Figure 1) provides an empirical index of state separability under the f-SNR model.

Extending this “decodability-as-fidelity” logic to large-scale connectomics, Guidotti et al. \cite{Guidotti2023} used fMRI functional connectivity patterns acquired during focused attention versus open monitoring to test whether meditation style leaves a discriminable signature—and whether this depends on expertise. Critically, meditation style was significantly decodable in expert Theravada Buddhist monks (mean accuracy 65.6\%), but not in novices who had only brief prior training (mean accuracy 54.4\%), with expert accuracies also significantly exceeding novice accuracies across folds. The discriminative features were concentrated in connectivity within/between the anterior salience network and dorsal default mode network, consistent with style-specific regulation of attention and self-related processing. In f-SNR terms, the key implication is the expertise-contingent emergence of a stable, separable neural geometry for distinct meditative modes—i.e., long-term practice appears to consolidate more reliable, lower-noise network-level representations of internal attentional states, making them easier to decode.

\subsection{A direct test of whether f-SNR increases with meditation depth}

\begin{figure}[htbp]
  \centering
  \caption{f-SNR measures across meditation depth}
  \label{fig:meditation_depth_fsnr}
  \includegraphics[width=0.95\textwidth]{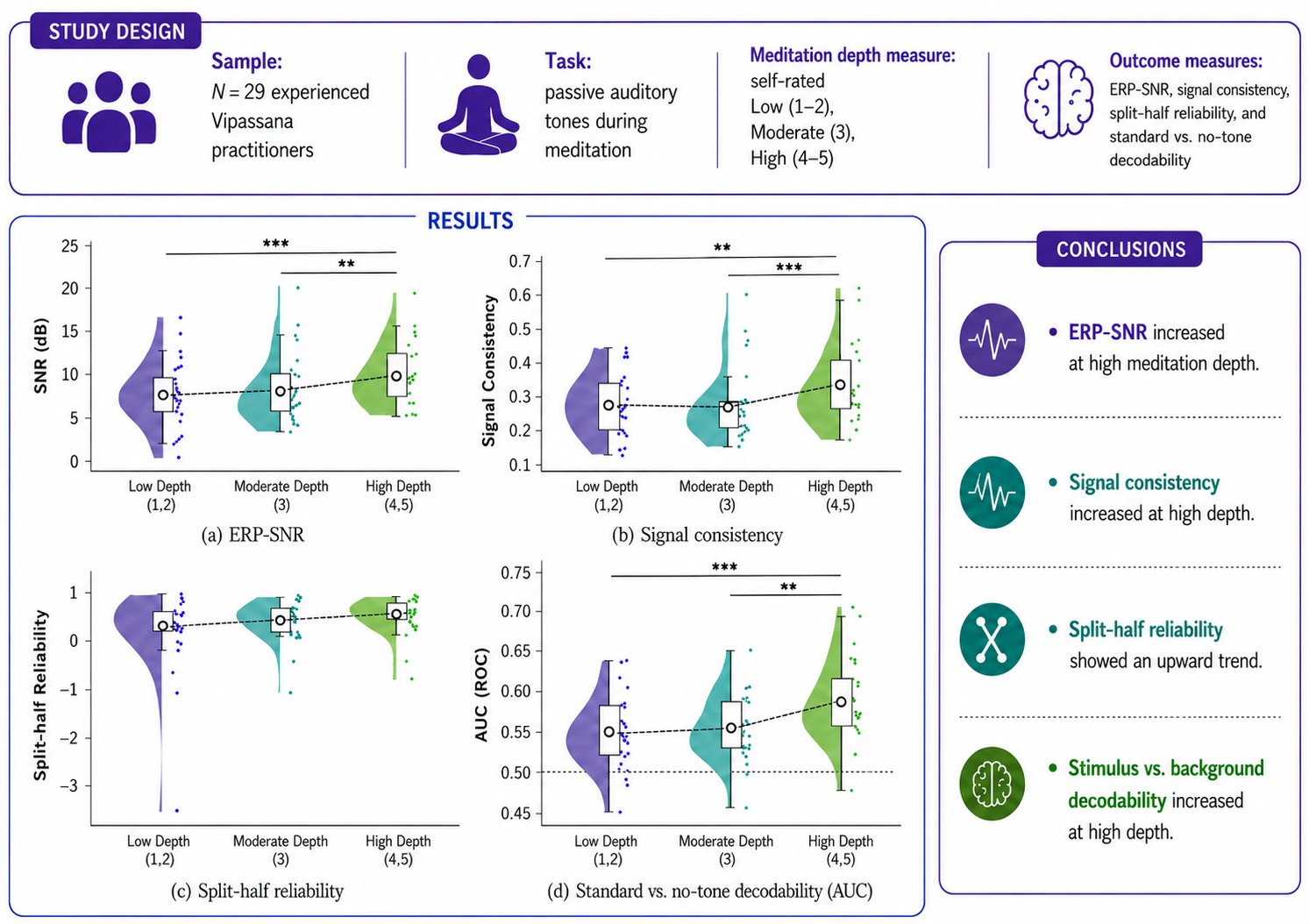}
  \begin{minipage}{0.95\textwidth}
  \small Note. Estimated values are shown for Low, Moderate, and High meditation depth for (A) ERP-SNR, (B) signal consistency, (C) split-half reliability, and (D) standard-versus-no-tone decodability. Asterisks indicate FDR-corrected pairwise comparisons: ${}^{**}p<.01$ and ${}^{***}p<.001$. The dotted line in panel (D) indicates chance-level classification performance ($\mathrm{AUC}=.50$). Adapted from Nath et al. \cite{Nath2026}.
  \end{minipage}
\end{figure}

A recent EEG study was conducted specifically as a direct test of the f-SNR meditation hypothesis: namely, that neural representations of sensory events should become clearer and more recoverable as meditation deepens \cite{Nath2026}. Twenty-nine experienced Vipassana practitioners meditated while a passive stream of auditory tones was presented using the Sustained Auditory Passive Oddball (SAPO) paradigm, periodically reporting the depth of their meditation. EEG segments associated with these reports were grouped into Low, Moderate, and High meditation depth. The study operationalised f-SNR using four complementary measures derived from responses to frequent standard tones: ERP-SNR, signal consistency, split-half reliability, and standard-versus-no-tone decodability (Figure~\ref{fig:meditation_depth_fsnr}).

High-depth meditation was associated with significantly greater ERP-SNR and signal consistency than both Low and Moderate depth. Standard-versus-no-tone decodability also increased at High depth, indicating that tone-evoked activity could be distinguished more accurately from ongoing background EEG. Split-half reliability increased numerically from Low to Moderate to High depth, although was slightly below the conventional significance threshold (p=0.058). By contrast, meditation depth did not reliably affect classification between the standard, oddball, and distractor auditory stimulus types. The depth-related effect therefore appeared to concern the separation of stimulus-evoked activity from background neural activity, rather than a general improvement in the discrimination of auditory events.

This pattern is particularly relevant to the proposed decluttering mechanism. The standard tones were frequent, predictable, task irrelevant, and required no overt response, making it unlikely that the results were driven primarily by deliberate attention to the stimuli. Instead, deeper meditation potentially reduced endogenous activity that would otherwise obscure the stimulus-locked response, allowing the same sensory event to be represented with greater consistency and separability. The study therefore provides initial within-subject empirical support for the f-SNR meditation hypothesis, and more specifically for the proposal that meditation can increase f-SNR by reducing endogenous interference. 

\subsection{Future work and a toolkit for falsification}

For f-SNR to serve as a meaningful lens on meditation or cognition more broadly, it must be empirically tractable and readily falsifiable. To this end, f-SNR can be indexed or inferred using many well-established techniques that quantify either the fidelity of neural activity relative to causes \cite{Borst1999,Knill2004}, or the reliability of neural activity associated with the same cause over time \cite{Zuo2014,Mathewson2015}. Table 2 provides a series of techniques for estimating the brain’s SNR and accompanying experiments and predictions.

Several key studies are yet to be conducted. Perhaps the most urgent study is one that specifically tracks the phenomenological quality of “clarity” alongside deepening meditation (e.g., on an extended retreat) while simultaneously measuring changes in key markers of f-SNR. If f-SNR is a reflection of how well the brain’s internal geometry mirrors the causal structure of its environment, then it provides a novel information-theoretic bridge between phenomenology and neurobiological activity, very much in the spirit of computational neurophenomenology \cite{SandvedSmith2025,Lutz2025,Ramstead2022,Laukkonen2025}. At the trait level, we would expect advanced meditators can voluntarily enter states of high f-SNR more easily and deeply than non-meditators or novice practitioners. Another clear test of the present theory is to compare mutual information between sensory data and neural activity, under the expectation that temporally extended mutual information between input streams and neural activity will be higher in advanced practitioners than novices (i.e., more closely tracking the signal manifold, Figure 1). Ultimately, convergent neurobiological measures of f-SNR could serve as an objective yardstick of meditation expertise (Table 2).

An important consideration for future work is the potential of U-shaped curves, wherein initial phases of meditation increase noise as the mind transitions from ordinary states to more stable ‘contemplative’ states. This is consistent with reports that meditation can be experienced initially as destabilising before settling into new, more “stable”, attractor basins or homeostatic setpoints \cite{Lindahl2017,Sparby2025}. Hence, careful comparisons across state, trait, and depth of meditation are needed, with concomitant tracking of phenomenological and neurobiological variables.

\begin{table}[htbp]
  \centering
  \footnotesize
  \caption{How to measure f-SNR: Methods and predictions}
  \label{tab:2}
  \begin{tabularx}{\textwidth}{>{\raggedright\arraybackslash}X>{\raggedright\arraybackslash}X>{\raggedright\arraybackslash}X}
    \toprule
    Measure of SNR & Example Methodology & Prediction / Interpretation \\
    \midrule
    Trial-to-trial variability (variability quenching) \cite{Arazi2019,Lutz2009} & EEG/MEG during visual/auditory oddball or attention task: compare variance of evoked responses before vs after meditation. & Meditation reduces trial-to-trial variability (enhanced reliability of evoked responses), indicating improved cortical stability and attentional gain control. \\
    Mutual information between stimuli and neural responses \cite{Ince2017} & Compute MI between visual stimuli and EEG channel / fMRI responses. & Higher MI in meditators reflects greater shared information (higher SNR) between external causes and neural representations. \\
    Cross-validated decoding accuracy (MVPA / RSA separability) \cite{Kriegeskorte2008,Pereira2009,Varoquaux2017} & Train classifier to decode presented stimuli (e.g., faces vs objects) from fMRI signals. & Greater decoding accuracy or more orthogonal representational manifolds in meditators implies higher representational SNR. \\
    Temporal SNR (tSNR) in fMRI \cite{Triantafyllou2011} & Compute voxelwise tSNR (mean / SD of time series). & Increased tSNR during meditation suggests reduced physiological and motion noise, more efficient large-scale coordination. \\
    TMS–EEG perturbation SNR \cite{She2024} & Deliver single-pulse TMS to parietal cortex and compute early TMS evoked potential (TEP) amplitude / noise ratio across trials. & Meditation yields higher TEP SNR (less trial-to-trial variability), indicating a more coherent cortical response regime. \\
    Reliability metrics (split-half / ICC) \cite{Mathewson2015,Groppe2009} & Compute split-half reliability of EEG alpha power during meditation vs rest. & Higher reliability across segments during meditation shows more stable oscillatory dynamics (reduced endogenous noise). \\
    ERP waveform similarity (single-trial to mean correlation / lagged cross-correlation) \cite{Gaspar2011,Li2009} & (i) Pearson correlation between each trial waveform and the grand-average (template) within a component-specific time window, and/or (ii) maximum cross-correlation within a small lag window to allow latency jitter. & Higher similarity in meditators indicates more consistent waveform morphology across trials, consistent with higher f-SNR. \\
    Criticality-based SNR (scaling exponents / avalanches) \cite{Yu2022} & Record MEG / fMRI / EEG; estimate power-law exponents and branching ratio & Activity moves toward critical regime (balanced excitation/inhibition), maximizing dynamic range and f-SNR in meditation (cf. Section 4). \\
    \bottomrule
  \end{tabularx}
\end{table}

\section{f-SNR, Criticality, and Computational Efficiency}

The drive toward a high f-SNR state may be rooted in a fundamental evolutionary pressure for metabolic efficiency. The human brain is a remarkably expensive organ, consuming about 20\% of the body’s energy budget while comprising only about 2\% of its mass \cite{HerculanoHouzel2011,Raichle2002}. Neural signaling itself is metabolically costly, with major energetic demands arising from synaptic transmission, action potentials, and the restoration of ionic gradients after neural activity \cite{Attwell2001}. This creates a powerful selective pressure to optimize the energy cost of neural activity. A noisy, cluttered mind, with incessant and/or redundant mental, emotional, and physiological noise, represents a state of high-cost, inefficient computation.

One powerful way that complex systems improve their efficiency is through sustaining a so-called “critical” or near-critical regime of activity: a state poised at a phase transition between order (subcritical) and chaos (supercritical). A growing body of evidence suggests that the brain, like many complex systems in nature, operates near a critical point \cite{Cocchi2017,OByrne2022,Toyoizumi2011}. This state of “self-organized criticality” is widely argued to confer functional advantages for information processing. In particular, operating near a critical point can maximize dynamic range (the span of input intensities a network can encode) \cite{Kinouchi2006} and is associated with peaks in information capacity and information transmission in cortical-network experiments \cite{Shew2013}, and fluctuations in cognitive performance \cite{Muller2025}. Interestingly, a growing body of research also shows that mutual information is maximised at criticality \cite{Shew2013} consistent with higher f-SNR. For the present framework, the criticality axis may be treated as a global “gain-setting” dimension that shapes how efficiently and flexibly a system models a changing world (i.e., its effective f-SNR) \cite{Kinouchi2006,OByrne2022}.

As illustrated in Figure 4, the present f-SNR framework can be deepened by proposing that meditation is a form of mental training that allows for the voluntary tuning of the brain's operating point along the critical-subcritical axis. The idea is that meditation provides a way to strike a balance between overly rigid precision (subcritical; under-responsive, habit-locked, low dynamic range) and runaway volatility (supercritical; hypersensitive, unstable, high internal noise). The result is a regime in which useful signals are amplified appropriately through a network, with less noise dominating global dynamics.

Indeed, recent studies provide convergent evidence that advanced meditation is associated with systematic shifts of brain dynamics toward near-critical regimes. For example, whole-brain modeling of advanced jhāna practitioners shows that progressively deeper absorption states are characterized by a collapse of heterogeneous network dynamics toward near-bifurcation working points \cite{Vohryzek2025}, particularly within the default mode network (DMN). Complementing these results, also during jhāna meditation, practitioners show increased signal diversity across multiple metrics (e.g., Lempel–Ziv complexity, entropy measures), coupled with reduced chaoticity and preserved or enhanced responsiveness to perturbation \cite{Mago2025}. The work of Dürschmid et al. \cite{Durschmid2020} aligns even more specifically with the f-SNR perspective, with MEG analyses showing that: “…avalanche dynamics shifted towards a critical point during meditation by reduction of neural noise”.

\begin{figure}[htbp]
  \centering
  \caption{Criticality applied to different organisms and meditation states}
  \label{fig:4}
  \includegraphics[width=0.95\textwidth]{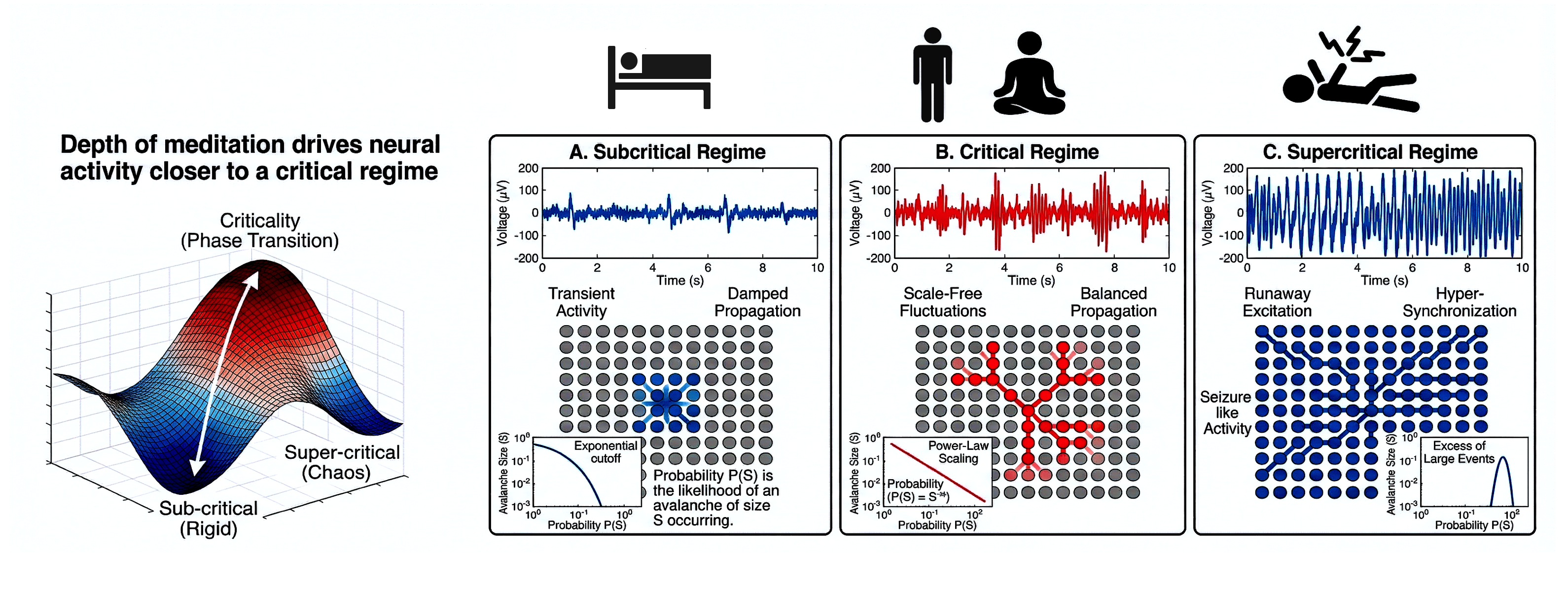}
  \begin{minipage}{0.95\textwidth}
  \small Note. The figure presents three distinct dynamical states of neural activity, with corresponding simulated EEG traces, schematic neuronal propagation patterns, and avalanche size probability distributions \cite{Vohryzek2025}. A) Subcritical Regime (e.g., non-REM sleep, general anaesthesia, or coma): The simulated EEG trace shows low-amplitude, rapidly damped fluctuations. The neuronal grid illustrates Transient Activity with Damped Propagation, where activity is localized to a small cluster of neurons. The log-log plot of avalanche size (S) versus probability (P(S)) exhibits an "Exponential cutoff," indicating that large neuronal avalanches are extremely rare. B) Near Critical Regime (e.g., ordinary < meditative mind): The EEG trace demonstrates balanced, scale-free fluctuations with a mix of small and large events, with meditation shifting the ordinary mind towards criticality. The grid shows Scale-Free Fluctuations and Balanced Propagation, characterized by complex, branching patterns of activity that are neither damped nor runaway. The avalanche size distribution follows a straight line on the log-log plot, indicating Power-Law Scaling (P(S) \textasciitilde{} $S^{-\alpha}$), a hallmark of criticality associated with optimal information processing and long-range correlations. C) Supercritical Regime (e.g., epileptic seizure): The EEG trace shows high-amplitude, sustained, and synchronized activity, resembling a seizure including runaway excitation and hyper-Synchronization, characteristic of epilepsy.
  \end{minipage}
\end{figure}

Crucially, we need to consider the potential for non-linear curves along the meditation journey, e.g., wherein beginner and advanced practices may differ in their effects on EEG temporal structure and criticality-related features \cite{Walter2022,Irrmischer2018}. Experiences of “awakening” or insight may also correspond to sudden but sustained increases in criticality. For example, evidence from rare meditative endpoints known as “cessations” further supports this account \cite{Chowdhury2023}. EEG recordings surrounding sudden gaps/cessations in consciousness associated with deep meditation reveal transient increases in long-range temporal correlations, with detrended fluctuation analysis exponents approaching values associated with criticality \cite{vanLutterveld2025}. Notably, these shifts begin prior to the loss of consciousness and persist afterward. Such findings align with practitioners’ reports of a profound “reset” or post-cessation clarity—possibly driven by a non-linear boost in criticality and f-SNR.

Taken together, it is plausible that meditation enhances the brain’s f-SNR not only through selective decluttering and amplification (cf. Figure 1) and regulated hierarchical engagement (cf. Figure 2), but also by globally through shifting the brain closer to a critical regime. This, in turn, raises the hypothesis that meditation makes a system more metabolically efficient over time. However, as yet, no agreed upon mechanism exists for how meditation shifts neural activity closer to criticality.

Two candidate mechanisms emerge from earlier discussion: 1) Mindfulness may optimise “branching activity patterns” in response to perturbation, analogous to the way that it regulates hierarchical engagement (cf. Section 3.1, Figure 2). This amounts to responding in a balanced way to the stream of data, not too much, not too little. 2) Deepening meditation (as in jhāna, insight practice, cessation, or minimal phenomenal experience) may further relax rigid priors (e.g., the self-model, see also \cite{Carter2005,DorZiderman2025}), permitting a more flexible responsiveness to the stream of incoming data; supporting critical-like neural dynamics.

\section{Implications for Psychopathology: Why Meditation is Panacea-Like}

Meditation is often criticized as a “panacea”: a single intervention claimed to help everything from anxiety to addiction \cite{VanDam2018}. While this criticism is warranted, it also points to a real empirical pattern: meditation can exert broad, cross-domain effects \cite{Goyal2014,Creswell2017}. The f-SNR framework offers a principled explanation for why meditation can appear panacea-like.

If f-SNR indexes the fidelity of the brain’s inference, then improving f-SNR should benefit any condition in which symptoms are maintained by noisy, unstable, or mis-weighted representations. In this sense, f-SNR functions as an upstream, transdiagnostic variable: shifting it even modestly might propagate improvements across attention, affect regulation, and belief updating, yielding wide but non-uniform clinical benefits \cite{Friston2023}.

Indeed, computational psychiatry often frames psychopathology as reflecting failures of selective enhancement and decluttering, i.e., maladaptive precision control \cite{Feldman2010,Friston2014}. For example, depression has been framed as excessive endogenous noise in the form of self-referential rumination, where repetitive internal priors dominate processing and reduce sensitivity to corrective evidence from ongoing experience \cite{NolenHoeksema2008}. Anxiety disorders can be framed as pathological amplification of threat-related signals (and/or threat priors), producing hypervigilance and narrow, self-reinforcing sampling of evidence \cite{BarHaim2007}. Obsessive compulsive disorders can be construed as over-precision of error, coupled to compulsive action policies that reduce uncertainty locally while preventing broader model revision \cite{Gillan2014,Fradkin2020}. ADHD also fits naturally as a disorder of low task-relevant f-SNR: weak or unstable amplification of goal signals and insufficient decluttering of distractors yields volatile control and high susceptibility to capture \cite{Hauser2016}. At the extreme end, delusions (and related psychotic symptoms) can be understood as low-fidelity inference due to aberrant salience—impairing calibration to present-moment data and trapping the system in self-confirming interpretations or unstable world-models \cite{Kapur2003,Adams2013,Sterzer2018}.

Crucially, this framing also clarifies why meditation is not a universal remedy. First, “signal” is goal- and context-relative: enhancing interoception, for example, may be therapeutic for some individuals and destabilizing for others \cite{Creswell2017}. Second, increasing sensitivity to internal data can transiently increase experienced noise, especially when long-avoided affective material becomes more available to awareness \cite{Lindahl2017}. Third, different practices tune different channels of precision (focused attention, open monitoring, constructive practices, and deconstructive/non-dual approaches)  \cite{Dahl2015}. The practical implication is that meditation’s broad efficacy is most plausibly explained not by disorder-specific mechanisms, but by a shared computational target—improving the fidelity of inference by titrating inferences to the ongoing stream of present-moment data \cite{Goyal2014,Friston2023}.

Beyond meditation, we may find that a convergent measure of f-SNR provides a uniquely predictive biomarker of various psychopathologies, wherein the severity of psychopathology negatively correlates with f-SNR. Conversely, positive states of human flourishing may yield a positive correlation. Clearly, the clinical application of the f-SNR framework is an exciting domain for future research.

\section{Implications for Technology}

Perhaps the most provocative implication of this framework is the fact that meditation may, for the first time, be integral for the functioning of emerging technologies. The f-SNR framework reframes meditation as a training regime that makes neural activity easier to measure, decode, and stabilize for technological coupling. This matters because many BCI bottlenecks are fundamentally SNR problems: non-invasive signals are weak, artifact-prone, and non-stationary, and decoders often require substantial calibration because neural features drift across time and differ markedly between individuals \cite{Khademi2023,Huang2021}.

If meditation reliably increases the separability, reliability, and stationarity of task-relevant neural patterns (i.e., boosts f-SNR), then it should reduce the “burden” placed on sensors and machine learning—shifting part of the performance problem from engineering to the brain itself. Indeed, recent work evinces this interpretation.

In a 12-week randomized intervention with an active control, mindfulness meditation training improved EEG-BCI accuracy more than music training or no-treatment control, suggesting effects beyond expectancy and generic engagement \cite{Tan2014}. Mechanistic work further indicates that mind–body awareness training can accelerate BCI learning and selectively increases controllable alpha-band activity during BCI control, with alpha dynamics predicting final performance and relating to practice intensity \cite{Stieger2021}. A large-scale longitudinal study similarly reported that mindfulness-based stress reduction improved SMR-based BCI performance, accompanied by increased frontolimbic alpha activity and altered alpha connectivity consistent with reduced processing of unnecessary information during control \cite{Jiang2021}. These findings fit naturally within the framework that meditation improves the f-SNR of the brain.

If meditation improves the clarity of information in the brain, this could substantially reduce calibration time, improve cross-day generalization, and expand what is feasible with low-cost non-invasive hardware, because the upstream neural manifold becomes cleaner and less entangled with mind-wandering, reactive appraisal, and self-referential noise. In parallel, closed-loop neurofeedback systems could be used to teach these states more efficiently—treating f-SNR as an explicit training target and titrating difficulty/dosage in real time \cite{Chen2025}. Over the longer term, this suggests a plausible pathway toward higher-bandwidth bidirectional interfaces that leverage the benefits of meditation or actively regulate the target of f-SNR directly, for example, via brain criticality.

\section{Conclusions}

This paper has mapped the diverse phenomenological and neurobiological landscape of meditation onto a single, empirically tractable coordinate: the brain’s functional signal-to-noise ratio. The quintessential notion of “mental clarity” is redefined as a quantifiable state where brain activity tightly tracks relevant causes rather than noise. By selectively decluttering distractions and enhancing signals (e.g., concentration), optimising hierarchical engagement (e.g., mindfulness), and shifting global dynamics toward a thermodynamically efficient critical regime (e.g., advanced practices), meditation may render the brain a more reliable modeler of its environment. The implications of this shift extend far beyond the cushion. For psychiatry, f-SNR offers a transdiagnostic target, reframing conditions from anxiety and ADHD to psychosis as extended failures of signal-to-noise inference that can be regulated through meditation or other f-SNR interventions. For emerging technology, this framework suggests a new design principle: clarify the brain to optimize the machine. As we approach a future of high-bandwidth human-computer interaction, the capacity to voluntarily regulate the fidelity of one’s own neural code may become a functional necessity for maintaining agency in a complex, AI-infused, world.

\section*{Acknowledgments}

I would like to acknowledge Mihir Nath, Evan Lewis-Healey, Edmundo Lopez-Sola, and Shamil Chandaria for their valuable comments and feedback on this work.

\nocite{*}
\printbibliography

\end{document}